\documentclass{article}
\usepackage{amsmath,amsthm}
\usepackage{amssymb,latexsym}
\usepackage[mathscr]{eucal}
\usepackage{setspace}
\usepackage{graphics}
\usepackage{array}

\newcommand{\al}{\alpha}
\newcommand{\na}{\nabla}
\newcommand{\pa}{\partial}
\newcommand{\pr}{\prime}

\newcommand{\eps}{\epsilon}
\newcommand{\lp}{\left(}
\newcommand{\rp}{\right)}
\newcommand{\lb}{\left[}
\newcommand{\rb}{\right]}
\newcommand{\lc}{\left\{}
\newcommand{\rc}{\right\}}
\newcommand{\la}{\langle}
\newcommand{\ra}{\rangle}
\newcommand{\be}{\begin{equation}}
\newcommand{\ee}{\end{equation}}

\newcommand{\dell}{d_e/\ell \ll 1}
\newcommand{\delg}{d_e/\ell \gg 1}
\newcommand{\delln}{d_e/\ell_n \ll 1}
\newcommand{\delgn}{d_e/\ell_n \gg 1}
\newcommand{\ihat}{\bf\hat{i}}

\newcommand{\done}{D_{0_{(1)}}}
\newcommand{\dtwo}{D_{0_{(2)}}}
\newcommand{\yal}{y_\al^2}
\newcommand{\kal}{k_\al^2}
\newcommand{\desq}{d_e^2}

\begin{document}

\begin{center}
\huge{\bf Electron Magnetohydrodynamic Turbulence: Universal Features}

\vspace{.3in}

\large{B.K. Shivamoggi\footnote{\normalsize Permanent Address: University of Central Florida, Orlando, FL 32816-1364}\\
Kavli Institute of Theoretical Physics,\\
University of California, Santa Barbara, CA 93106}
%Telephone: 407-823-2061\\
%Fax: 407-823-6253\\
%E-mail: bhimsen.shivamoggi@ucf.edu}
\end{center}

\vspace{.3in}

\begin{flushright}
\end{flushright}

\noindent\Large{\bf Abstract}\\

\large The energy cascade of electron magnetohydrodynamic (EMHD) turbulence is considered. Fractal and multi-fractal models for the energy dissipation field are used to determine the spatial intermittency corrections to the scaling behavior in the high-wavenumber (electron hydrodynamic limit) and low-wavenumber (magnetization limit) asymptotic regimes of the inertial range. Extrapolation of the multi-fractal scaling down to the dissipative microscales confirms in these asymptotic regimes a {\it dissipative anomaly} previously indicated by the numerical simulations of EMHD turbulence. Several basic features of the EMHD turbulent system are found to be universal which seem to transcend the existence of the characteristic length scale $d_e$ (which is the electron skin depth) in the EMHD problem---

\begin{itemize}
\item {\it equipartition} spectrum,
\item Reynolds-number scaling of the dissipative microscales,
\item scaling of the probability distribution function (PDF) of the electron-flow velocity (or magnetic field) gradient (even with intermittency corrections),
\item {\it dissipative anomaly},
\item {\it critical exponent} scaling. 
\end{itemize}
\pagebreak

\noindent\Large{\bf 1. Introduction}\\

\large The high-temperature plasmas in space (e.g. solar flares and magnetospheric substorms) and laboratory (tokamak discharges) have been found to be collisionless. An important aspect of a collisionless plasma is the enhancement by an order of magnitude of the magnetic reconnection rate (Yamada \cite{Yam}). In situations where the spatial scales are shorter than the ion-inertial length $d_i$ and time scales are shorter than the ion-cyclotron period,\footnote{{\it In situ} measurements in the solar wind have provided evidence of magnetic field fluctuations characterized by such time (and spatial) scales (Alexandrova et al. \cite{Ale}).} the ions do not have time to respond and merely provide a neutralizing background, and the dynamics are controlled entirely by electrons. A fluid description for the electrons then leads to the electron magnetohydrodynamic (EMHD) model (Kingsep et al. \cite{Kin}, Gordeev et al. \cite{Gor}). EMHD, unlike MHD, has a characteristic length scale $d_e$ (the electron inertial length) which turns out to control the strength of nonlinearity in EMHD. The strongly sheared electron flows in the current sheets in EMHD undergo Kelvin-Helmholtz instability and lead to turbulence in EMHD (which is to be contrasted with turbulence generation/intensification via the tearing mode instability of current sheets in MHD). The energy cascade in EMHD turbulence proceeds directly even in two dimensions (2D), as in MHD turbulence, thanks to the Lorentz force on the electrons. Biskamp et al. \cite{Bis}, \cite{Bis2} did high resolution numerical simulation of decaying 2D isotropic homogeneous EMHD turbulence and found that the energy spectrum follows the Kolmogorov spectrum in the electron hydrodynamic limit ($\delg$) in spite of the fact that the whistler waves (which are generic to EMHD) would be expected to mediate the energy cascade. (A whistler-like relation\footnote{The extent of the whistlerization in EMHD turbulence was numerically investigated by Dastgeer et al. \cite{Das}.} implying an equipartition of energy between the poloidal and axial components of the magnetic field was however found to hold.) Celani et al. \cite{Cel} further showed that a Kolmogorov 4/5th law type result also holds for the energy cascade in 2D EMHD turbulence. Numerical simulations of Boffetta \cite{Bof} revealed the presence of spatial intermittency in EMHD turbulence - the energy dissipation field was found not to be uniformly distributed in space and the dissipative structures were of filament shape. Numerical simulations of Germaschewski and Grauer \cite{Ger} showed deviations from a Kolmogorov-type linear law of the characteristic scaling exponent of higher order structure functions further validating this aspect. Numerical simulations of Biskamp et al. \cite{Bis} and \cite{Bis2} also showed that the energy dissipation rate in EMHD turbulence was apparently independent of the dissipation coefficients suggesting the possibility of a {\it dissipative anomaly}\footnote{The finiteness of the energy dissipation even in the limit the dissipation coefficients vanish constitutes a {\it dissipative anomaly} (persistence of symmetry breaking even in the limit the symmetry breaking factors vanish). There is experimental support (Sreenivasan \cite{Sre}) for this in 3D hydrodynamic turbulence.} in the direct energy cascade in EMHD. 

In this paper, we consider fractal (Frisch et al. \cite{Fri}) and multi-fractal (Frisch and Parisi \cite{Fri2}) models\footnote{{\it  In situ} measurements in a stationary interval of fast ambient solar wind have provided evidence of plasma turbulence showing a cross over from being multi-fractal in the inertial range to being fractally homogeneous in the dissipative range (Kiyani et al \cite{Kiy}).} to describe the effects of spatial intermittency in 2D fully-developed EMHD turbulence. We will then extrapolate multi-fractal scaling in the inertial range down to the dissipative microscale and provide analytical evidence for a \textit{dissipative anomaly} in the high-wavenumber (electron hydrodynamic limit) and low-wavenumber (magnetization limit) asymptotic regimes. Several basic features of the EMHD turbulent system which seem to be universal and transcend the existence of the characteristic length scale $d_e$ in the EMHD problem are highlighted.

\vspace{.3in}

\noindent\Large{\bf 2. Governing Equations of EMHD}\\

\large The 2D EMHD system of equations can be written in terms of two scalar potentials - the magnetic flux function $A$ describing the in-plane magnetic field ${\bf B} = \na \times A ~\ihat_z$ and the stream function $\psi$ describing the in-plane electron flow velocity in the plane ${\bf v}_e = \na \times \psi ~\ihat_z$, which is proportional to the in-plane current density (so $\psi$ also represents the out-of-plane magnetic field):\\

\begin{itemize}
  \item the equation of generalized vorticity:
\be\tag{1}
\frac{\pa}{\pa t} \lp \omega + \frac{\psi}{\desq} \rp + \lp {\bf v}_e \cdot \na \rp \omega - \frac{1}{m_e n_e c} \lp {\bf B} \cdot \na \rp J = \frac{\nu}{\desq} \na^2 \omega
\ee
  \item the generalized Ohm's law:
\be\tag{2}
\frac{\pa}{\pa t} \lp A + \frac{\desq}{c} J \rp + \lp {\bf v}_e \cdot \na \rp \lp A + \frac{\desq}{c} J \rp = \eta \na^2 A
\ee
\end{itemize}
where,
\be\tag{3}
\frac{1}{c} J = -\na^2 A, ~\omega = -\na^2 \psi,
\ee
and $\eta$ is the resistivity and $\nu$ is the kinematic viscosity of the plasma.

The number density $n_e$ is constant, in accordance with the incompressibility of the electron flow $\na \cdot {\bf v}_e = 0$ which implies $\na \cdot {\bf J} = 0$ - this presupposes that the displacement current $\pa {\bf E}/\pa t$ is negligible.

In the ideal limit ($\nu$ and $\eta \Rightarrow 0$), equations (1) and (2) have the Hamiltonian integral invariant (upon appropriately non-dimensionalizing the various quantities (Biskamp et al. \cite{Bis} and \cite{Bis2})),
\be\tag{4}
H = \frac{1}{2} \iint\limits_S \lb \lp \na A \rp^2 + \psi^2 + \desq \lc J^2 + \lp \na \psi \rp^2 \rc \rb d S
\ee
S being the area occupied by the plasma. (4) shows that the dissipation effects introduce a characteristic length scale, namely, $d_e$ in the EMHD problem, which turns out to control the strength of the nonlinearity in EMHD. As a result, the latter exhibits some departures from the properties of MHD turbulence. One such feature is a decrease of the energy flux, leading to energy pileup of scales $\ell_n \sim d_e$ in the energy cascade. This could lead to an ordered quasi-crystalline phase signifying the appearance of long-range order in the system (similar to the case with geostrophic turbulence (Kukharin et al. \cite{Kuk}) and kinetic Alfv\'{e}n turbulence (Shivamoggi \cite{Shi})).

(4) implies, on noting a whistler-like relation\footnote{This relation also implies an equipartition in the energy contents of the in-plane magnetic field and velocity fluctuations (Alexandrova et al. \cite{Ale}).} $\psi \sim A/\ell$ holds between the poloidal and axial components of the magnetic field (Biskamp et al. \cite{Bis} and \cite{Bis2}), that the energy per unit mass at length scale $\ell$ is given by
\be\tag{5}
E \sim \psi^2 \lp 1 + \frac{\desq}{\ell^2} \rp
\ee
which, in the magnetization $(\dell)$ and the electron hydrodynamic $(\delg)$ asymptotic regimes, leads to
\be\tag{6a, b}
E \sim \lc
\begin{matrix}
\psi^2, ~\dell\\
\lp \desq/\ell^2 \rp \psi^2, ~\delg.
\end{matrix}
\right.
\ee

It is of interest to note that EMHD turbulence also exhibits some basic features which transcend the existence of the characteristic length scale $d_e$ in the EMHD problem. One such feature becomes apparent on applying the equilibrium statistical mechanics approach to the EMHD problem.

\vspace{.3in}

\noindent\Large{\bf 3. Equilibrium Statistical Mechanics}\\

\large Consider an EMHD turbulence within a square which can be expanded into an infinite series of discrete wave vectors ${\bf k}_n$ with stream function amplitudes $\Psi ({\bf k}_n)$ related to each other via equations (1) and (2). The Fourier analysis of this system allows a formulation in terms of many degrees of freedom and hence leads to a consideration of this problem from the viewpoint of statistical mechanics.

Application of equilibrium statistical mechanics to this system (Burgers \cite{Bur}, Hopf \cite{Hop}, Lee \cite{Lee} and Kraichnan \cite{Kra}) requires the latter to be considered ideal.\footnote{Formally, equilibrium statistical mechanics does not seem to be applicable to turbulence which, being dissipational, is in a non-equilibrium state. However, a turbulent system is believed to relax via nonlinear interactions toward equilibrium (which was confirmed for the 3D hydrodynamics case by the numerical calculations of Orszag and Patterson \cite{Ors}). Indeed, one may interpret the energy cascade to small length scales as a consequence of this tendency (Novikov \cite{Nov}).} This, in turn, requires a truncation in the Fourier space by discarding the Fourier modes higher than a cut-off wavenumber $k_{\text{max}}$. This truncated set of N wavenumbers conserves the energy (according to (5)), which is a quadratic rugged invariant (because it is conserved by an interacting triad),
\be\tag{7}
\frac{1}{2} \sum\limits_{{\bf k}_n} \lp 1 + k^2_n \desq \rp | \Psi \lp {\bf k}_n \rp |^2 = const.
\ee

If $y_{n 1} \lp {\bf k}_n \rp$ and $y_{n 2} \lp {\bf k}_n \rp$ are the real and imaginary parts of each mode $\Psi \lp {\bf k}_n \rp$, the system can be represented by a point of $m \equiv 2N$ coordinates in a phase space and evolves ergodically in this phase space on the energy sphere,
\be\tag{8}
\frac{1}{2} \sum\limits_{\al = 1}^m \lp 1 + \kal \desq \rp \yal = const.
\ee

Consider now a collection of such systems which is represented at each instant of time by a cluster of points in the phase space of density $\rho \lp y_1, ..., y_m, t \rp$. Since the total number of such systems and hence the volumes occupied by their representative points in the phase space are preserved, we have the {\it Liouville Theorem}:
\be\tag{9}
\frac{\pa \rho}{\pa t} + \sum\limits_{\al = 1}^m \frac{d y_\al}{d t} \frac{\pa \rho}{\pa y_\al} = 0.
\ee

Statistical mechanics seeks to explain the statistical behavior of a system in terms of its structural properties, such as the conservation of energy. This enables the equilibrium spectrum of EMHD turbulence to be predicted from the viewpoint of canonical ensemble averages.

The equilibrium solutions of Liouville's equation (9) may be constructed as functions of the conserved quantities, such as the energy (8), via the {\it Boltzmann-type distribution},
\be\tag{10}
P \lp y_1, ..., y_m \rp = \frac{1}{Z} e^{-\frac{1}{2} \sum\limits_{\al = 1}^m \sigma \lp 1 + \kal \desq \rp \yal}
\ee
where $\sigma$ is a constant (interpretable as ``{\it inverse temperature}") and $Z$ is the partition function of the system,
\be\tag{11}
Z \equiv \int ... \int e^{-\frac{1}{2} \sum\limits_{\al = 1}^m \sigma \lp 1 + \kal \desq \rp \yal} d y_1 .... d y_m.
\ee
The canonical ensemble average $\la \rho \lp y_1, ..., y_m, t \rp \ra$ of an ensemble of the given system\\ $\rho \lp y_1, ..., y_m, t \rp$ is stipulated to relax eventually toward this equilibrium distribution over the energy sphere (8) in the phase space.

The mean variance of the mode $\al$ is given by
\be\tag{12a}
\la \yal \ra = \frac{1}{Z} \int ... \int \yal e^{-\frac{1}{2} \sum\limits_{\al = 1}^m \sigma \lp 1 + \kal \desq \rp \yal} d y_1 ... d y_m
\ee
or
\be\tag{12b}
\la \yal \ra = \frac{1/\sigma}{1 + \kal \desq}.
\ee

The energy spectrum is then given by
\be\tag{13a}
E (k) \sim \pi k \lp 1 + k^2 \desq \rp \la | \Psi ({\bf k}) |^2 \ra \sim \pi k, ~\forall k,
\ee
and using (12b),
\be\tag{13b}
E (k) \sim \pi k, ~\forall k.
\ee
(13b) shows that EMHD turbulence, like 2D hydrodynamic turbulence, exhibits the {\it equipartition} spectrum,\footnote{The number of modes in 2D is proportional to $2 \pi k$.} $E (k) \sim k$, for small wavenumbers. This result signifies basic dynamical aspects of EMHD turbulence which transcend the existence of the characteristic length scale $d_e$ in the EMHD problem, as is also apparent in the following developments.

\pagebreak
%\vspace{.3in}

\noindent\Large{\bf 4. Inertial-Range Scaling Laws}\\

\large One may consider for the energy cascade in EMHD turbulence an inertial range of Kolmogorov type which is in a state of statistical equilibrium and the energy is assumed to cascade smoothly through nonlinear processes in a stationary state.

Consider a discrete sequence of scales,
\be\tag{14}
\ell_n \sim \ell_0 \cdot 2^{-n} ~; ~n = 0, 1, 2, \cdots.
\ee
Let us assume that we have a statistically stationary EMHD turbulence, where energy is introduced into the plasma at scales $\sim \ell_0$, and is then transferred successively to smaller and smaller scales $\sim \ell_1, \ell_2, \cdots$ until some scale $\ell_d$ is reached where dissipative effects are able to compete with nonlinear transfer.

The energy per unit mass in the $nth$ scale, according to (5), is given by
\be\tag{15}
E_n \sim \Psi_n^2 \lp 1 + \frac{\desq}{\ell_n^2} \rp.
\ee

The rate of energy transfer per unit mass from the $nth$ scale to the $(n + 1)th$ scale is given by
\be\tag{16}
\eps_n \sim \frac{E_n}{t_n} \sim \frac{\Psi_n^3 d_e}{\ell_n^2} \lp 1 + \frac{\desq}{\ell_n^2} \rp
\ee
where $t_n$ is a characteristic time of the $nth$ scale,
\be\tag{17}
t_n \sim \frac{\ell_n^2}{d_e \Psi_n}.
\ee

In the inertial range, we assume a stationary process in which the energy transfer rate is constant,
\be\tag{18}
\eps_n = const = \eps, ~\ell_d \leq \ell_n \leq \ell_0.
\ee

Using (18), (16) leads to
\be\tag{19}
\Psi_n \sim \frac{\eps^{1/3} \ell_n^{2/3}}{d_e^{1/3}} \lp 1 + \frac{\desq}{\ell_n^2} \rp^{-1/3}.
\ee

Using (19), (15) gives
\be\tag{20}
E_n \sim \eps^{2/3} \frac{\ell_n^{4/3}}{d_e^{2/3}} \lp 1 + \frac{\desq}{\ell_n^2} \rp^{1/3}
\ee
from which we have
\be\tag{21a, b}
E_n \sim \lc
\begin{matrix}
\eps^{2/3} d_e^{-2/3} \ell_n^{4/3}, ~\delln\\
\eps^{2/3} \ell_n^{2/3}, ~\delgn.
\end{matrix}
\right.
\ee
(21) leads to the following energy spectra (Biskamp et al. \cite{Bis} and \cite{Bis2}),
\be\tag{22a, b}
E_k \sim \lc
\begin{matrix}
\eps^{2/3} d_e^{-2/3} k^{-7/3}, ~k d_e \ll 1\\
\eps^{2/3} k^{-5/3}, ~k d_e \gg 1.
\end{matrix}
\right.
\ee

The electron hydrodynamic limit corresponds to $k d_e \gg 1$ while the magnetization limit corresponds to $k d_e \ll 1$. The steeper energy spectrum, as per (22a, b), in the latter limit, signifies a weaker nonlinearity in the EMHD dynamics (governed by equations (1) and (2)) in this limit, as confirmed below in Section 5.

\pagebreak

\noindent\Large{\bf 5. Finite-Time Singularity in the Magnetic Field}\\

\large The existence of strongly localized features like current sheets in the small-scale structure of EMHD turbulence implies the development of singularities in the magnetic field. In order to see this, first note that the energy dissipation rate, as per (6), is given by
\be\tag{23a, b}
\eps \sim \lc
\begin{matrix}
\eta \displaystyle{\frac{\Psi^2}{\xi^2_{D_{(1)}}}}, ~\dell\\
\\
\nu \displaystyle{\frac{d_e^2\Psi^2}{\xi^4_{D_{(2)}}}}, ~\delg
\end{matrix}
\right.
\ee
$\xi_D$ being the dissipative microscale.

On using (19), (23a, b) becomes
\be\tag{24a, b}
\eps \sim \lc
\begin{matrix}
\eta \displaystyle{\frac{\eps^{2/3}d_e^{-2/3}}{\xi^{2/3}_{D_{(1)}}}}, ~\dell\\
\\
\nu \displaystyle{\frac{\eps^{2/3}}{\xi^{4/3}_{D_{(2)}}}}, ~\delg
\end{matrix}
\right.
\ee
from which,
\be\tag{25a, b}
\left.
\begin{matrix}
\xi_{D_{(1)}} \sim \displaystyle{\frac{\eta^{3/2}d_e^{-1}}{\eps^{1/2}}}, ~\dell\\
\\
\xi_{D_{(2)}} \sim \displaystyle{\frac{\nu^{3/4}}{\eps^{1/4}}}, ~\delg.
\end{matrix}
\rc
\ee

One may write for the evolution of the current density (in the magnetization limit \\$d_e/l\ll 1)$ and the electron vorticity (in the electron hydrodynamic limit $d_e/l\gg 1)$,
\be\tag{26a, b}
\left.
\begin{matrix}
\displaystyle{\frac{dJ}{dt}} \sim \eta \displaystyle{\frac{J}{\xi^2_{D_{(1)}}}}, ~\dell\\
\\
\displaystyle\frac{d \Omega} {dt} \sim \nu \displaystyle{\frac{\Omega}{\xi^2_{D_{(2)}}}}, ~\delg.
\end{matrix}
\rc
\ee

Using (25a, b), (26a, b) become
\be\tag{27a, b}
\left.
\begin{matrix}
\displaystyle{\frac{dJ}{dt}} \sim \displaystyle{\frac{\eps d_e^2}{\eta^2}} J, ~\dell\\
\\
\displaystyle\frac{d \Omega}{dt} \sim \displaystyle{\frac{\eps^{1/2}}{\nu^{1/2}}} \Omega, ~\delg.
\end{matrix}
\rc
\ee

On the other hand, a {\it dissipative anomaly} (confirmed below in Section 6.3) in EMHD turbulence, implies
\be\tag{28}
\eps \sim \lc
\begin{matrix}
\eta J^2, ~\dell\\
\\
\nu \Omega^2, ~\delg
\end{matrix}
\sim const.
\right.
\ee

Using (28), (27a, b) become
\be\tag{29a, b}
\left.
\begin{matrix}
\displaystyle{\frac{dJ}{dt}} \sim \displaystyle{\frac{d_e^2}{\eps}} J^5, ~\dell\\
\\
\displaystyle\frac{d \Omega}{dt} \sim \Omega^2, ~\delg
\end{matrix}
\rc
\ee
which imply,
\be\tag{30a, b}
\left.
\begin{matrix}
J \sim \displaystyle{\frac{1}{\lp t+C_1 \rp^{1/4}}}, ~\dell\\
\\
\Omega \sim \displaystyle{\frac{1}{\lp t + C_2 \rp}}, ~\delg
\end{matrix}
\rc
\ee
exhibiting finite-time singularities (FTS) in the magnetic field (in the magnetization limit $\dell$) and the electron velocity field (in the electron hydrodynamic limit $\delg$); $C_1$ and $C_2$ are arbitrary constants. Observe that the FTS in the magnetization limit is weaker than that in the electron hydrodynamic limit in agreement with a steeper energy spectrum, as per (22a,b).

\vspace{.3in}

\noindent\Large{\bf 6. Spatial Intermittency}\\

\large The Kolmogorov type inertial range theory discussed in Section 4 does not take into account the spatial intermittency in EMHD turbulence that was revealed by the numerical simulations (Boffetta et al. \cite{Bof} and Germaschewski and Grauer \cite{Ger}). Spatial intermittency effects would cause fluctuations in the energy dissipation rates and hence would lead to systematic departures from the scaling laws (22) which use mean transfer rates. The {\it global} statistical scaling invariance assumed by the Kolmogorov type inertial range theory is broken down by the spatial intermittency. However, one may still assume that the scaling invariance remains valid, nonetheless, {\it locally}. One may follow Mandelbrot \cite{Man} and argue that the spatial intermittency effects in EMHD turbulence are related to the fractal nature of the strongly convoluted dissipative structures (like the current sheets revealed in the numerical simulations \cite{Ger}). This may be simulated in a first approximation by representing the dissipative structures via a homogeneous fractal with non-integer Hausdorff dimension $D_0$. This amounts to assuming the energy flux to be transferred to only a fixed fraction $\beta$ of the eddies downstream in the cascade (Frisch et al. \cite{Fri}).

\vspace{.3in}

\noindent{\bf 6.1 Homogeneous Fractal Model}\\

We now assume that at the $nth$ step of the cascade, only a fraction $\beta^n$ of the total space with a fractal dimension $D_0$ has an appreciable excitation.

The energy per unit mass in the $nth$ scale is given by
\be\tag{31}
E_n \sim \beta^n \Psi_n^2 \lp 1 + \frac{\desq}{\ell_n^2} \rp
\ee
where,
\be\tag{32}
\beta^n \sim \lp \frac{\ell_n}{\ell_0} \rp^{2 - D_0}.
\ee

The energy transfer rate per unit mass from the $nth$ scale to the $(n + 1)th$ scale is given by
\be\tag{33}
\eps_n \sim \frac{E_n}{t_n} \sim \beta^n \frac{\Psi_n^3 d_e}{\ell_n^2} \lp 1 + \frac{\desq}{\ell_n^2} \rp.
\ee

In the inertial range, the energy transfer rate is constant for a stationary process, so on using (32) and (33), and assuming the scaling behavior,
\be\tag{34}
\Psi_n \sim \ell_n^\al
\ee
we have, from (18),
\be\tag{35a, b}
\left.
\begin{matrix}
3 \al + 2 - \done - 2 = 0, ~\delln\\
3 \al + 2 - \dtwo - 4 = 0, ~\delgn
\end{matrix}
\rc
\ee
from which, the H\"{o}lder scaling exponent $\alpha$ is given by,
\be\tag{36a, b}
\al = \lc
\begin{matrix}
\frac{\done}{3}, ~\delln\\
\frac{\dtwo + 2}{3}, ~\delgn.
\end{matrix}
\right.
\ee

Using (32), (34) and (36), we have from (31),
\be\tag{37a, b}
E_n \lp \ell_n \rp \sim \lc
\begin{matrix}
\eps^{2/3} d_e^{-2/3} \ell_n^{4/3 + 1/3 \lp 2 - \done \rp}, ~\delln\\
\eps^{2/3} \ell_n^{2/3 + 1/3 \lp 2 - \dtwo \rp}, ~\delgn.
\end{matrix}
\right.
\ee
(37) leads to the following energy spectra,
\be\tag{38a, b}
E (k) \sim \lc
\begin{matrix}
\eps^{2/3} d_e^{-2/3} k^{-7/3 - 1/3 \lp 2 - \done \rp}, ~k d_e \ll 1\\
\eps^{2/3} k^{-5/3 - 1/3 \lp 2 - \dtwo \rp}, ~k d_e \gg 1.
\end{matrix}
\right.
\ee
Observe that the intermittency corrections $\lp D_{0_{(1), (2)}} < 2 \rp$ make the spectra steeper, as expected.

Noting that in the electron hydrodynamic limit $(k d_e \gg 1)$ the dissipative structures are typically vortex-filament like $\lp \dtwo = 0 \rp$, and in the magnetization limit $(k d_e \ll 1)$ they are typically current-sheet like $\lp \done = 1 \rp$ (\cite{Ger}), (38a, b) would lead to
\be\tag{39a, b}
E (k) \sim \lc
\begin{matrix}
\eps^{2/3} d_e^{-2/3} k^{-8/3}, ~k d_e \ll 1\\
\eps^{2/3} k^{-7/3}, ~k d_e \gg 1.
\end{matrix}
\right.
\ee

On the other hand, noting that the structure function $S_p (\ell)$, of order $p$, for the EMHD turbulence problem is defined in terms of the magnetic field in the magnetization limit $(\dell)$ and the electron flow velocity in the electron hydrodynamic limit $(\delg)$, we have
\be\tag{40a, b}
S_p (\ell) \sim \lc
\begin{matrix}
\la |\delta \psi (\ell)|^p \ra, ~\dell\\
\la |\delta \lp \pa \psi/\pa \ell \rp (\ell)|^p, ~\delg.
\end{matrix}
\right.
\ee

Using (34), and noting that the probability to belong to this fractal at scale $\ell$ goes like $\ell^{2 - D_0}$, (40) leads to
\be\tag{41a, b}
S_p (\ell) \sim \ell^{\zeta_p} \sim \lc
\begin{matrix}
\ell^{\al p + 2 - \done}, ~\dell\\
\ell^{\lp \al - 1 \rp p + 2 - \dtwo}, ~\delg.
\end{matrix}
\right.
\ee
So, the characteristic exponent $\zeta_p$ is given by
\be\tag{42a, b}
\zeta_p = \lc
\begin{matrix}
\al p + 2 - \done, ~\dell\\
\lp \al - 1 \rp p + 2 - \dtwo, ~\delg.
\end{matrix}
\right.
\ee
Using (36), (42) becomes
\be\tag{43a, b}
\zeta_p = \lc
\begin{matrix}
\displaystyle{\frac{2p}{3}} - \lp \displaystyle{\frac{p}{3}} - 1 \rp \lp 2 - \done \rp, ~\dell\\
\\
\displaystyle{\frac{p}{3}} - \lp \displaystyle{\frac{p}{3}} - 1 \rp \lp 2 - \dtwo \rp, ~\delg
\end{matrix}
\right.
\ee
which does not show a nonlinear dependence on $p$, as is required of the characteristic exponent for large $p$. It is therefore necessary to consider the multi-fractal model (Frisch and Parisi \cite{Fri2}) to address this issue.

\vspace{.3in}

\noindent{\bf 6.2 Multi-fractal Model}\\

Let us assume that the energy flux (or dissipation) is concentrated on a multi-fractal object (\cite{Fri2}) which is characterized by a continuous spectrum of H\"{o}lder scaling exponents $\al ~, ~\al \in I \equiv \lb \al_{min}, \al_{max} \rb$. Each $\al \in I$ has the support set $S (\al) \subset \mathbb{R}^3$ of fractal dimension $f (\al)$ such that, as $\ell \Rightarrow 0$, the stream function increment has the scaling behavior\footnote{This is tantamount to assuming that the EMHD turbulence system possesses only a {\it local} scaling invariance, with the H\"{o}lder scaling exponent $\alpha$ varying from point to point in space.},
\be\tag{44}
|\delta \psi (\ell)| \sim \ell^{\al}.
\ee

The sets $S(\al)$ are nested so that $S \lp \al^\pr \rp \subset S (\al)$, for $\al^\pr < \al$. The fractal dimension $f(\al)$ is obtained via a Legendre transformation of the scaling exponent of the pth order structure function of the magnetic field (or electron-flow velocity),
\be\tag{45a, b}
S_p (\ell) \sim \lc
\begin{matrix}
\int d \mu (\al) \ell^{\al p + 2 - f(\al)} \sim \ell^{\zeta_{p_{(1)}}}, ~\dell\\
\int d \mu (\al) \ell^{(\al - 1) p + 2 - f(\al)} \sim \ell^{\zeta_{p_{(2)}}}, ~\delg
\end{matrix}
\right.
\ee
where the measure $d \mu (\al)$ gives the weight of different scaling exponents $\al$, and $\ell^{2 - f (\al)}$ represents the probability of encountering the set $S (\al)$ within a 2D circle of radius $\ell$. (45a, b) reflect the asymptotic scalings exhibited by (6a, b).

One may use the method of steepest descent to extract the dominant terms in the integrals in (45), in the limit of very small $\ell$. This gives
\be\tag{46a, b}
\zeta_p = \lc
\begin{matrix}
\al^* p + 2 - f(\al^*), ~d_e/\ell\\
\lp \al^* - 1 \rp p + 2 - f(\al^*), ~\delg
\end{matrix}
\right.
\ee
where,
\be\tag{46c}
f^\pr (\al_*) = p.
\ee

Next, in order to relate the singularity spectrum $f (\al)$ to the generalized fractal dimension (GFD) of the energy dissipation field, note that the energy transfer rate per unit mass at length scale $\ell$ is given by
\be\tag{47a, b}
\eps (\ell) \sim \frac{E \lp \ell \rp}{t \lp \ell \rp} \sim \lc
\begin{matrix}
\lp d_e/\ell^2 \rp \psi^3, ~\dell\\
\lp d_e^3/\ell^4 \rp \psi^3, ~\delg.
\end{matrix}
\right.
\ee

If the energy dissipation field is assumed to be a multi-fractal, the sums of the moments of the total energy dissipation $U (\ell) \sim \eps (\ell) \ell^2$ occurring in $N (\ell)$ squares of size $\ell$ covering the support of the measure $\eps$ exhibit the following asymptotic scaling behavior (Halsey et al. \cite{Hal})
\be\tag{48a, b}
\sum_{i = 1}^{N (\ell)} \lb U_i (\ell) \rb^q \sim \ell^{(q - 1) D_q} \sim \lc
\begin{matrix}
\int d \mu (\al) \ell^{3 \al q - f(\al)}, ~\dell\\
\int d \mu (\al) \ell^{(3 \al - 2) q - f(\al)}, ~\delg
\end{matrix}
\right.
\ee
where $D_q$ is the GFD of the $\eps$-field (Hentschel and Proccacia \cite{Hen}), and we have assumed that the number of {\it iso-}$\al$ squares for which $\al$ takes on values between $\al$ and $\al + d \al$ is proportional to $d \mu (\al) \ell^{-f (\al)}$. (48a, b) again reflect the asymptotic scalings exhibited by (6a, b). The dominant terms in the integrals in (48) may again be extracted, in the limit $\ell \Rightarrow 0$, using the method of steepest descent, to give
\be\tag{49a, b}
(q - 1) D_q = \lc
\begin{matrix}
3 \al^* q - f(\al^*), ~\dell\\
(3 \al^* - 2) q - f(\al^*), ~\delg
\end{matrix}
\right.
\ee
where,
\be\tag{49c}
f^\pr (\al^*) = 3q.
\ee
The coincidence of the values of $\al^*$ given by (46c) and (49c), for which the integrands in (45a, b) and (48a, b) become extremum, is insured by assuming a Kolmogorov refined similarity type hypothesis (Meneveau and Sreenivasan \cite{Men}) in the dissipative microscale regime.

Eliminating $f(\al)$ from (46) and (49), and putting $q = p/3$, we obtain
\be\tag{50a, b}
\zeta_p = \lc
\begin{matrix}
\displaystyle{\frac{2p}{3}} - \lp \displaystyle{\frac{p}{3}} - 1 \rp \lp 2 - D_{p/3} \rp, ~\dell\\
\\
\displaystyle{\frac{p}{3}} - \lp \displaystyle{\frac{p}{3}} - 1 \rp \lp 2 - D_{p/3} \rp, ~\delg.
\end{matrix}
\right.
\ee

For a fractally homogeneous EMHD turbulence,
\be\tag{51a, b}
D_{p/3} = \lc
\begin{matrix}
\done, ~\dell\\
\dtwo, ~\delg
\end{matrix}
\right.
~\forall p
\ee
(50a,b) reduce to
\be\tag{52a, b}
\zeta_p = \lc
\begin{matrix}
\displaystyle{\frac{2p}{3}} - \lp \displaystyle{\frac{p}{3}} - 1 \rp \lp 2 - \done \rp, ~\dell\\
\\
\displaystyle{\frac{p}{3}} - \lp \displaystyle{\frac{p}{3}} - 1 \rp \lp 2 - \dtwo \rp, ~\delg
\end{matrix}
\right.
\ee
in agreement with (43a, b). The energy per unit mass then shows the following scaling behavior,
\be\tag{53a, b}
E (\ell) \sim \lc
\begin{matrix}
\eps^{2/3} d_e^{-2/3} \ell^{4/3 + 1/3 \lp 2 - D_{0_{(1)}} \rp}, ~\dell\\
\eps^{2/3} \ell^{2/3 + 1/3 \lp 2 - D_{0_{(2)}}\rp}, ~\delg
\end{matrix}
\right.
\ee
and the energy spectra are,
\be\tag{54a, b}
E (k) \sim \lc
\begin{matrix}
\eps^{2/3} d_e^{-2/3} k^{-7/3 - 1/3 \lp 2 - D_{0_{(1)}} \rp}, ~k d_e \ll 1\\
\eps^{2/3} k^{-5/3 - 1/3 \lp 2 - D_{0_{(2)}} \rp}, ~k d_e \gg 1
\end{matrix}
\right.
\ee
in agreement with (38a, b).

\vspace{.3in}

\noindent{\bf 6.3 Multi-fractal Scaling at the Dissipative Microscale}\\

We now consider extrapolation of the multi-fractal scaling in the inertial range discussed in Section 6.2 down to the dissipative microscale by assuming that an inertial behavior persists at scales smaller than $d_e$ - this assumption may be justifiable for tenuous plasmas like those in space ($d_e \approx$ 10 km for the magnetospheric plasma).

On taking into account the spatial intermittent character of the energy dissipation field, the dissipative microscales $\xi_{D_{(1), (2)}}$, given by (25a, b), (along the lines of the development of Paladin and Vulpiani \cite{Pal}, Sreenivasan and Meneveau \cite{SM}, and Nelkin \cite{Nel} for the hydrodynamic case), are found to exhibit the scaling behavior,
\be\tag{55a}
\xi_{D_{(1)}} \sim \bar{R}_m^{-1/\al}, ~\dell
\ee
\be\tag{55b}
\xi_{D_{(2)}} \sim \bar{R}_h^{-1/\al}, ~\delg
\ee
where $\bar{R}_m$ and $\bar{R}_h$ are, respectively, mean magnetic and hydrodynamic Reynolds numbers,
\be\tag{56}
\bar{R}_m \sim \frac{\lp \bar{\eps} \ell^5/d_e \rp^{1/3}}{\eta}, ~\bar{R}_h \sim \frac{\lp \bar{\eps} \ell^7/d_e^3 \rp^{1/3}}{\nu}
\ee
$\bar{\eps}$ is the mean energy dissipation rate. (Observe that the mean magnetic and hydrodynamic Reynolds numbers are both dependent on the electron skin depth $d_e$.) The identity of the scaling exponents in the two opposite asymptotic regimes, as indicated by (55a, b), is symptomatic of certain universal features in these regimes, as seen further in the following.

The moments of the magnetic field (or electron-flow velocity) -gradient distribution,
\be\tag{57a, b}
A_p \equiv \lc
\begin{matrix}
\la | \pa \psi/\pa x |^p \ra, ~\dell\\
\la | \pa^2 \psi/\pa x^2 |^p \ra, ~\delg
\end{matrix}
\right.
\ee
are then given by
\be\tag{58a, b}
A_p \sim \lc
\begin{matrix}
\int d \mu (\al) \lp \bar{R}_m \rp^{-\frac{1}{\al} \lb (\al - 1) p + 2 - f(\al) \rb}, ~\dell\\
\int d \mu (\al) \lp \bar{R}_h \rp^{-\frac{1}{\al} \lb (\al - 2) p + 2 - f(\al) \rb}, ~\delg.
\end{matrix}
\right.
\ee

In the limit of large $\bar{R}_m$ and $\bar{R}_h$, the dominant exponents in (58) correspond to
\be\tag{59a}
\al^* \lb p - f^\pr (\al^*) \rb = (\al^* - 1) p + 2 - f(\al^*), ~\dell
\ee
\be\tag{59b}
\al^* \lb p - f^\pr (\al^*) \rb = (\al^* - 2) p + 2 - f(\al^*), ~\delg.
\ee
The coincidence of the values of $\al^*$ given by (49c) and (59a, b) for which the integrands in (48a, b) and (58a, b) become extremum, is again insured by assuming the Kolmogorov refined similarity type hypothesis (\cite{Men}) in the dissipative microscale regime. (59a, b), in conjunction with (49a, b), lead to
\be\tag{60a, b}
A_p \sim \lc
\begin{matrix}
(\bar{R}_m)^{-\displaystyle{\frac{D_Q (p - 3) - 3p + 6}{D_Q}}}, ~\text{where $Q$ is the root of} &Q = \displaystyle{\frac{D_Q + p - 2}{D_Q}}, ~\dell\\
(\bar{R}_h)^{-\displaystyle{\frac{D_Q (p - 3) - 6p + 10}{D_Q ~+ ~2}}}, ~\text{where $Q$ is the root of} &Q = \displaystyle{\frac{D_Q + 2p - 2}{D_Q + 2}}, ~\delg.
\end{matrix}
\right.
\ee
Here, the detailed dependence of $D_Q$ on $Q$ is not required for the present discussion. We have from (60a, b),
\be\tag{61a, b}
A_2 \sim \lc
\begin{matrix}
(\bar{R}_m)^1, ~\dell\\
(\bar{R}_h)^1, ~\delg.
\end{matrix}
\right.
\ee
So, the mean energy dissipation has the following scaling behavior,
\be\tag{62a, b}
\left.
\begin{matrix}
\eta A_2 \sim (\bar{R}_m)^0, ~\dell\\
\\
\nu A_2 \sim (\bar{R}_h)^0, ~\delg.
\end{matrix}
\rc
\ee
(62a, b) implies an inviscid dissipation of energy in the electron hydrodynamic limit and a non-resistive dissipation of energy in the magnetization limit and hence a {\it dissipative anomaly} in high- and low-wavenumber asymptotic regimes of EMHD turbulence in confirmity with DNS (\cite{Bis}, \cite{Bis2}). Note further from (60a, b) that the energy dissipation field in these asymptotic regimes has the GFD $D_Q$ equal to the information entropy dimension $D_1$. It is of interest to note that these results hold in both the high- and low-wavenumber asymptotic regimes of EMHD turbulence in spite of the disparate strength of the nonlinearity in EMHD in these two asymptotic regimes (as shown in Sections 4 and 5). The {\it dissipative anomaly} therefore signifies another basic dynamical aspect of EMHD turbulence which transcends the existence of the characteristic length $d_e$ in the EMHD problem.

\vspace{.3in}

\noindent\Large{\bf 7. Discussion}\\

\large One may view the energy dissipation rate $\eps$ to be the order parameter {\it \'{a} la} Landau \cite{Lan} for the EMHD turbulence problem because it appears to indicate the degree of broken symmetry and exhibits fluctuations in the presence of spatial intermittency. Further, noting that the critical point for EMHD turbulence corresponds to the limit $\bar{R}_m$ and $\bar{R}_h \Rightarrow \infty$, the non-zero limiting value of $\eps$, as the critical point is approached appears to validate this view.\footnote{This perspective therefore allows (Shivamoggi \cite{Shi2}) the Kolmogorov type inertial range theory described in Section 4, which assumes $\eps$ to be uniform, to be appropriately regarded as a mean field theory (MFT) {\it \'{a} la} Landau \cite{Lan}. The MFT neglects the spatial fluctuations in the order parameter $\epsilon$ (which become very important near the critical point), and hence does a poor job in describing the behavior of the system near the critical point, as to be expected. Thus, the {\it global} statistical scaling invariance assumed by the Kolmogorov type inertial range theory breaks down near the critical point ({\it spontaneous symmetry breaking}), and the scaling invariance may be assumed to be {\it local} in this region ({\it symmetry subgroup invariance}).} Indeed, one may define the critical exponent $\sigma$\footnote{One of the goals of critical phenomena formulation of the turbulence problem (Nelkin \cite{Nel2}, Yakhot and Orszag \cite{Yak}, Eyink and Goldenfeld \cite{Eyi}, Esser and Grossmann \cite{Ess}, Shivamoggi \cite{Shi2}) has been to determine the critical exponents that are intrinsic features of the turbulence dynamics and are not artifacts of the large-scale turbulence generation mechanisms.} (Shivamoggi \cite{Shi2}) for this problem by

\be\tag{63}
\eps \sim \lc
\begin{matrix}
\lp \bar{R}_m \rp^\sigma, ~\bar{R}_m \Rightarrow \infty, ~\dell\\
\lp \bar{R}_h \rp^\sigma, ~\bar{R}_h \Rightarrow \infty, ~\delg
\end{matrix}
\right.
\ee
where, as per (60a, b),
\be\tag{64}
\sigma = 3 (Q - 1), ~\forall d_e/\ell.
\ee

Comparison of the above results (see Table 1) with the corresponding results for various FDT systems in fluid and plasma dynamics (Shivamoggi \cite{Shi3} - \cite{Shi5}) indicates that the energy (or enstrophy in 2D hydrodynamic FDT) dissipation rate $\eps$ is the right choice for the order parameter for the FDT problem,\footnote{It may be mentioned that different choices (Nelkin \cite{Nel2}, Rose and Sulem \cite{Ros}) have been considered for the order parameter for the FDT problem; the present choice seems to be appealing because it agrees with all the implications posited in Landau's order parameter concept (see also footnote 10, as well as remark above equation (63)).} with an apparently {\it universal} form for the critical exponent $\sigma$ given by\footnote{The critical exponent $\sigma$ may be connected with the critical exponents $\gamma$, $\nu$ and $\eta$ introduced by Rose and Sulem \cite{Ros}, according to
\be\notag
\left.
\begin{matrix}
k_d \equiv \xi_D^{-1} \sim \bar{R}^\nu \sim \bar{R}^{1/\al} ~\text{so} ~\nu = 1/\al\\
S_2 (\ell) \sim \ell^\eta \sim \ell^{\zeta_2} ~\text{so} ~\eta = \zeta_2
\end{matrix}
\rc
\ee
as follows,
\be\notag
\sigma = 3 \lp Q - 1 \rp
\ee
with
\be\notag
Q = \frac{\gamma + 2}{3}, \gamma = \nu \lp 2 - \eta \rp.
\ee
Noting, from (36) and (37), that
\be\notag
\nu = \lc
\begin{matrix}
3/2 ~, ~d_e/\ell \ll 1\\
3/4 ~, ~d_e/\ell \gg 1
\end{matrix}
\right.
\ee
and
\be\notag
\eta = \lc
\begin{matrix}
4/3 ~, ~d_e/\ell \ll 1\\
2/3 ~, ~d_e/\ell \gg 1
\end{matrix}
\right.
\ee
we obtain
\be\notag
\left.
\begin{matrix}
Q = \gamma = 1\\
\sigma = 0
\end{matrix}
\rc \forall ~d_e/\ell
\ee
as required.}
\be\tag{65}
\sigma = a (Q - 1).
\ee
The variations in the amplitude $a$ reflect the residual effect of variant cascade physics in the diverse FDT systems.\footnote{This is totally in accord with the idea of universality which implies that near a critical point all systems can be grouped into a relatively small number of classes (depending on the specific dynamics) with identical critical exponents within each class (Hohenberg and Halperin \cite{Hoh}).} Observe in Table 1 that the energy (or enstrophy) dissipation fields in the various FDT systems have the GFD $D_Q$ equal to the information entropy dimension $D_1$ (because, corresponding to $p = 2$, the GFD index $Q$ turns out to be unity for all these FDT cases).

\begin{table}
   \begin{center}
     \begin{tabular}{|c|c|c|}
       \hline\\
       FDT case & Critical Exponent $\sigma$ & Generalized Fractal Dimension Index \\[.3in] \hline\\
       3D incompressible FDT & $3 (Q - 1)$ & $Q = \displaystyle{\frac{D_Q + 2p - 3}{D_Q + 1}}$ \\[.3in] \hline\\
       2D incompressible \\FDT-enstrophy cascade & $3 (Q - 1)$ & $Q = \displaystyle{\frac{D_Q + 3p - 2}{D_Q + 4}}$ \\[.3in] \hline\\
       3D compressible FDT & $\lp \displaystyle{\frac{3 \gamma - 1}{\gamma + 1}} \rp (Q - 1)$ & $Q = \displaystyle{\frac{D_Q + \displaystyle{\frac{2 \gamma}{\gamma + 1}} p - 3}{D_Q + \displaystyle{\frac{4 \gamma}{\gamma + 1}} - 3}}$ \\[.3in] \hline\\
       3D MHD FDT & $2 (Q - 1)$ & $Q = \displaystyle{\frac{2D_Q + 3p - 6}{2D_Q}}$ \\[.3in] \hline\\
       2D EMHD FDT & $3 (Q - 1)$ & $Q = \lc \begin{matrix} \displaystyle{\frac{D_Q + p - 2}{D_Q}}, ~\dell\\ \displaystyle{\frac{D_Q + 2p - 2}{D_Q + 2}}, ~\delg \end{matrix} \right.$ \\[.3in] \hline
     \end{tabular}
     \caption{Critical exponents for various FDT cases.}
   \end{center}
\end{table}

Further insight can be gained into this aspect by looking at the probability distribution function (PDF) of the electron-flow velocity (or magnetic field) -gradient. In order to derive the PDF of the magnetic field (or electron-flow velocity) -gradient, note that the scaling behavior of the dissipative microscales, on using (25a, b), is given by

\be\tag{66a, b}
\left.
\begin{matrix}
\xi_{D_{(1)}} \sim \lp \displaystyle{\frac{\eta}{\psi_0}} \rp^{1/\al}, ~\dell\\
\\
\xi_{D_{(2)}} \sim \lp \displaystyle{\frac{\nu}{\psi_0}} \rp^{1/\al}, ~\delg
\end{matrix}
\rc
\ee
where $\psi_0$ is the stream function increment on a macroscopic length L.

The scaling behavior of the magnetic field (or electron-flow velocity) -gradient is then
\be\tag{67a, b}
s \sim \lc
\begin{matrix}
\displaystyle{\frac{\psi}{\xi_{D_{(1)}}}} \sim \psi_{\displaystyle 0}^{{1/\al}} \eta^{\lp \al - 1 \rp/\al}, ~\dell\\
\displaystyle{\frac{d_e \psi}{\xi_{D_{(2)}}^2}} \sim d_e \psi_{\displaystyle 0}^{2/\al} \nu^{\lp \al - 2 \rp/\al}, ~\delg.
\end{matrix}
\right.
\ee

The PDF of the magnetic field (or electron-flow velocity) -gradient may then be determined in terms of that for the characteristic stream function increment $\psi_0$ for large scales as follows,
\be\tag{68}
P (s; \al) = P (\psi_0) \frac{d \psi_0}{d s}.
\ee
Taking $P (\psi_0)$ to be Gaussian,
\be\tag{69}
P (\psi_0) \sim e^{-\psi_0^2/2 <\psi_0^2>}
\ee
and using (67a, b), (68) leads to
\be\tag{70a, b}
P (s; \al) \sim \lc
\begin{matrix}
\lp \displaystyle{\frac{\eta}{|s|}} \rp^{1 - \al} ~e^{-\lb \displaystyle{\frac{\eta^{2 (1 - \al)} |s|^{2 \al}}{2 <\psi_0^2>}} \rb}, ~\dell\\
\lp \displaystyle{\frac{\nu}{|s|^{1/2}}} \rp^{2 - \al} ~e^{-\lb \displaystyle{\frac{\nu^{2 (2 - \al)} |s|^{\al}}{2 <\psi_0^2>}} \rb}, ~\delg.
\end{matrix}
\right.
\ee

For EMHD turbulence, on noting from (36),
\be\tag{71a, b}
\al = \lc
\begin{matrix}
2/3, ~\dell\\
4/3, ~\delg
\end{matrix}
\right.
\ee
(70a, b) become
\be\tag{72a, b}
P (s) \sim \lc
\begin{matrix}
\lp \displaystyle{\frac{\eta}{|s|}} \rp^{1/3} ~e^{-\lb \displaystyle{\frac{\eta^{2/3} |s|^{4/3}}{2 <\psi_0^2>}}\rb}, ~\dell\\
\lp \displaystyle{\frac{\nu^2}{|s|}} \rp^{1/3} ~e^{-\lb \displaystyle{\frac{\nu^{4/3} |s|^{4/3}}{2 <\psi_0^2>}}\rb}, ~\delg.
\end{matrix}
\right.
\ee

The identity of the $|s|$-dependence exhibited by $P (s)$, as per (72a, b), (which is also the same as the PDF for the velocity gradient for 3D hydrodynamic turbulence given by Frisch and She \cite{Fri3}), appears to be consistent with the indication of {\it dissipative anomaly}, as per (62a, b), in the asymptotic regimes (this is validated further by the critical exponent (64) for EMHD). The stretched exponential decay of the PDF exhibited by (72) has also been indicated by the numerical simulations (\cite{Bof} and \cite{Ger}).\footnote{Upon incorporating the intermittency corrections as per the homogeneous fractal model, given by (36), (72) becomes
\be\notag
P (s) \sim \lc
\begin{matrix}
\lp \frac{\eta}{|s|} \rp^{\lp 1 - \done/3 \rp} e^{-\lb \frac{\eta^{2 \lp 1 - \done/3 \rp} |s|^{2 \done/3}}{2 < \psi_0^2 >} \rb}, ~\frac{d_e}{\ell} \ll 1\\
\\
\lp \frac{\nu}{|s|^{1/2}} \rp^{\lp 4 - \dtwo\rp/3} e^{-\lb \frac{\nu^{2 \lp 4 - \dtwo\rp/3} |s|^{\lp\dtwo + 2\rp/3}}{2 < \psi_0^2 >} \rb}, ~\frac{d_e}{\ell} \gg 1.
\end{matrix}
\right.
\ee
Noting that in the electron hydrodynamic limit $\lp \delg \rp$ the dissipative structures are typically vortex-filament like $\lp \dtwo = 0 \rp$, and in the magnetization limit $\lp \dell \rp$ they are typically current-sheet like $\lp \done = 1 \rp$ (\cite{Ger}), the above expressions lead to
\be\notag
P (s) \sim \lc
\begin{matrix}
\lp \frac{\eta}{|s|} \rp^{2/3} e^{-\lb \frac{\eta^{4/3} |s|^{2/3}}{2 < \psi_0^2 >} \rb}, ~\frac{d_e}{\ell} \ll 1\\
\\
\lp \frac{\nu^2}{|s|} \rp^{2/3} e^{-\lb \frac{\nu^{8/3} |s|^{2/3}}{2 < \psi_0^2 >} \rb}, ~\frac{d_e}{\ell} \gg 1.
\end{matrix}
\right.
\ee
The identity of the $|s|$ - dependence exhibited by $P (s)$ in both the asymptotic regimes $\lp \dell ~\text{and} ~\delg \rp$ is again consistent with the indication of dissipative anomaly in these regimes. Observe the enhanced exponential stretching of the PDF due to intermittency, as to be expected.} 

\vspace{.3in}

\noindent\Large{\bf Acknowledgments}\\

\large I am thankful to Professors K.R. Sreenivasan and S.C. Chapman for their helpful comments and Dr. A. Das for helpful discussions. This research was supported in part by NSF grant No. PHY05-51164.

\end{document}